\documentclass[11pt,twoside]{article}
\usepackage{newpasp}
\usepackage{epsf}
\usepackage{graphics}

\newcommand{\chandra}{{\em Chandra}}
\newcommand{\ccd}{CCD}
\newcommand{\psf}{PSF}

\newcommand{\acis}{ACIS}

\begin{document}

\title {Five Years of Observations with the \chandra\ X-Ray Observatory\\
}

\author{M.C.Weisskopf}
\affil{NASA/MSFC, SD50, MSFC AL 35812}

\begin{abstract}
The \chandra\ X-ray Observatory is the X-ray component of NASA's Great Observatory Program and has been operating successfully for over five years. 
We present here brief overview of the technical performance and some of the remarkable discoveries.
\end{abstract}

\section{Technical Performance}
\subsection{The Spacecraft}
The Observatory  was launched on July 23, 1999 using the Space Shuttle Columbia. Placement in its highly elliptical orbit was completed 15 days after launch.  
The orbit has a nominal apogee of 140,000 km and a nominal perigee of 10,000 km. 
With this orbit, the satellite is above the radiation belts for more than about
75\% of the 63.5-hour orbital period and uninterrupted observations lasting more than 2 days are possible.
The observing efficiency, which also depends on solar activity, has varied from
65\% to more than 70\%. 

The spacecraft has been functioning superbly since launch.  
The specified design life of the mission was 5 years; however, the only
perishable is gas for maneuvering and is sized to allow operation for much more than 10 years. 
The orbit will be stable for decades. 

One of the most important functions of the Observatory subsytems is to provide the absolute accuracy of \chandra\ X-ray source locations.  
Based on observations of 225 point sources detected within 2\arcmin\ of the
boresight and having accurately known coordinates, the 90\% source location
error circle has a radius of 0.64\arcsec\ and fewer than 1\% of sources are outside a 1\arcsec\ radius.  

There is also an effective blurring of the X-ray point spread function (\psf) due to uncertainties in determining the aspect solution.
An indirect method of estimating aspect blurring uses the aspect solution to ``de-dither'' the aspect camera's star images and measure the residual jitter.  
Based on a sample of 350 observations one finds that 99\% of the time the
effective blurring is less than 0.20\arcsec\ (FWHM).

\subsection{The Optics}
The heart of the Observatory is the X-ray telescope made of four concentric,
precision-figured, superpolished Wolter-1 telescopes, similar to those
used for both the Einstein and Rosat observatories, but of much higher quality,
larger diameter, and longer focal length. 

The telescope's on-axis \psf, as measured during ground calibration,
had a full-width at half-maximum less than 0.5 arcsec and a half-power
diameter less than 1 arcsec. 
The pre-launch prediction for the on-orbit encircled-energy fraction was that a
1-arcsec-diameter circle would enclose at least half the flux from a point
source. 
A relatively mild dependence on energy, resulting from diffractive scattering by
surface microroughness, attests to the better than 3-angstroms-rms surface
roughness measured with optical metrology during fabrication and confirmed by
the ground X-ray testing.
The on-orbit performance met expectations.

\subsection{The Focal Plane Cameras \label{s:cameras}}

The Pennsylvania State University (PSU, University Park, Pennsylvania) and MIT
designed and fabricated the Advanced CCD Imaging Spectrometer (ACIS) with CCDs produced by MIT's Lincoln Laboratory. 
Made of a 2-by-2 array of large-format, front-illuminated (FI), 2.5-cm-square
CCDs, ACIS-I provides high-resolution spectrometric imaging over a
17-arcmin-square field of view.
ACIS-S, a 6-by-1 array of 4 FI CCDs and two back-illuminated (BI) CCDs mounted
along the \chandra\ transmission grating dispersion direction, serves both as the primary read-out detector for the High Energy Transmission Grating (HETG),
and, using the one BI CCD which can be placed at the aimpoint of the telescope,
also provides high-resolution spectrometric imaging extending to lower energies
but over a smaller (8-arcmin-square) field than ACIS-I.
Both ACIS detectors are covered with aluminized-polyimide filters, designed to
block visible light.
The spatial resolution for imaging with \acis~is limited by the
physical size of the \ccd~pixels (24.0 ${\mu}$m square $\sim$0.492
arcsec). 

The ACIS FI CCDs originally approached the theoretical limit for energy
resolution at almost all energies, while the BI were of somewhat lesser quality
due to imperfections induced in the manufacturing process (Bautz et al. 1998).
Subsequent to launch and orbital activation, the FI CCDs have developed
much larger charge transfer inefficiency (CTI) and the energy resolution has become a function of the row number, being nearer pre-launch values close to the frame store region and progressively degraded toward the farthest row (Prigozhin et al.2000). Note that a post-facto software correction has been developed which recovers much of the energy resolution lost (Townsley et al. 2002).

The damage to the FI CCDs was caused by low energy protons which Rutherford-scattered from the X-ray telescope
onto the focal plane.
Subsequent to the discovery of the degradation, operational procedures were
changed so that the ACIS instrument is not left at the focal position during
radiation belt passages where the proton flux is the highest.
Since then, degradation in performance has been limited
to the small, gradual increase due to cosmic rays that was predicted before
launch.
The BI CCDs were not impacted as it
is far more difficult for low energy protons to deposit their energy in the
buried channels where damage is most detrimental to performance.
These channels are near the CCD gates and the BI gates face in the direction
opposite to the telescope.
  
The ACIS instrument has the coldest surfaces on the Observatory. 
Thus, naturally, there has been a gradual accumulation of a contaminating layer
which decreases the low energy detection efficiency and introduces distinctive absorption edges.
The contaminant appears to be deposited on the optical blocking filters and
not on the CCDs themselves.
These filters are nominally at a temperature of -60 degrees C.   

An obvious response to contaminant deposition is to temporarily raise the
temperature of the focal plane.  
This procedure has not been immediately undertaken because of the experience,
early in the mission, in which a bakeout further degraded the CTI of the
already radiation-damaged FI CCDs.  
Currently the risks and benefits of different bakeout scenarios are under
study.

The Smithsonian Astrophysical Observatory (SAO, Cambridge MA),  designed and
fabricated the other focal plane camera the High Resolution Imager (HRC) (Murray et al. 2000).
Made of a single 10-cm-square microchannel plate (MCP), the HRC-I provides
high-resolution imaging over a 30-arcmin-square field of view.
A second detector made of 3 rectangular MCP segments (3-cm $\times$ 10-cm each) mounted end-to-end along the grating dispersion direction, the HRC-S, serves as the primary read-out detector for the Low Energy Transmission Grating (LETG).
Both detectors have cesium-iodide-coated photocathodes and are covered with
aluminized-polyimide UV/ion shields.
There has been no noticeable buildup of contamination on these filters which are much warmer than the ACIS filters. 

The HRC time resolution of 16 $\mu$sec offers the highest precision timing of
the two imaging cameras.
However, the HRC was mis-wired so that the time of the event associated with the
j-th trigger is that of the previous (j-th -1) trigger. 
If the all the data were routinely telemetered, the times could be simply corrected by reassigning the time tag. 
Since the problem has been discovered, new operating modes have been defined
which allow one to telemeter all data whenever the total counting rate is
moderate to low. 
For very bright sources the counting rate is so high that information associated
with certain triggers are never telemetered. 
In this case, the principal reason for dropping events is that the on-board,
first-in-first-out (FIFO) buffer fills as the source is introducing events at a
rate faster than the telemetry readout. 
Events are dropped until readout commences freeing one or more slots in the
FIFO. 
This situation can also be dealt with (Tennant et al. 2001) at the price of reduced detection efficiency and time resolution of the order of a millisecond can be achieved.

\subsection{The Gratings}

Aft of the X-ray telescope are 2 objective transmission gratings (OTGs) -
the Low-Energy Transmission Grating (LETG) and the High-Energy Transmission
Grating (HETG).
Positioning mechanisms are used to insert either OTG into the converging beam
where they disperse the x-radiation onto the focal plane.

The Space Research Institute of the Netherlands and the Max-Planck-Instit\"ut f\"ur extraterrestrische Physik designed and fabricated the LETG.
When in position, the 540 grating facets, mounted 3 per module, lie tangent to the Rowland toroid which includes the focal plane.  
With free-standing gold bars of about 991-nm period, the LETG provides
high-resolution spectroscopy ($E/\Delta E$ $>1000$) between 80 and 175 \AA\/
(0.07 -- 0.15 keV) and moderate resolving power at shorter wavelengths.  
The nominal LETG~wavelength range accessible with the HRC-S as the detector is
1.2 -- 175 \AA~(0.07 -- 10 keV); ACIS-S coverage is 1.2 -- 65 \AA~(0.20 -- 10
keV).
The on-orbit performance of the LETG is similar to pre-flight predictions
(e.g.\ Brinkman et al.\ 1997; Predehl et al.\ 1997; Dewey et al.\
1998), 

The dominant contribution to the LETG line response function (LRF) and
instrument resolving power is the telescope \psf, which is $\sim25~\mu$m FWHM,
depending on energy.  
When the LETG is used with the HRC-S, the intrinsic uncertainty in photon
position determination adds another small contribution of order 15-$20~\mu$m.
Uncertainties in correcting photon event positions and for the observatory
aspect also introduces an additional blurring of order a few $\mu$m.
For spectral lines with $\la 1000$ counts, the FWHM of the combined LETG+HRC LRF can is $\sim40$~$\mu$m, or $\sim 0.05$~\AA.  

The Massachusetts Institute of Technology (MIT, Cambridge, Massachusetts)
designed and fabricated the HETG.
The HETG employs 2 types of grating facets~--- the Medium-Energy Gratings (MEG),
which, when swung into position appear behind the X-ray telescope's 2 outermost shells, and the High-Energy Gratings (HEG), which appear behind the X-ray telescope's 2 innermost shells.
With polyimide-supported gold bars of 400-nm and 200-nm periods, respectively,
the HETG provides high-resolution spectroscopy from 0.4 to 4 keV (MEG, 30 to 3
\AA) and from 0.8 to 8 keV (HEG, 15 to 1.5 \AA).

The HETGS LRF has a Gaussian-like core with extended wings. 
The model of the HETG LRF is comprised of two Gaussians and two Lorentzians with
the narrow Gaussian dominating.
The LRFs derived from the fits match in-flight data extremely well (Marshall, Dewey, \& Ishibashi, 2004).
The spectral resolution is the FWHM of the LRF and is, for all practical purposes, independent of wavelength being 0.012\AA\ for the HEG and 0.023\AA\ for the MEG. 

\section{Observations and Discoveries \label{s:discoveries}} 

\begin{figure}
\begin{center} 
\epsfysize=7cm
\epsfbox{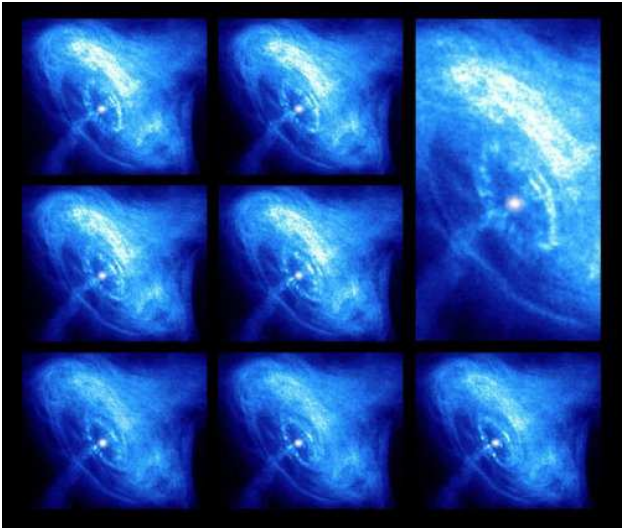} 
\caption{Eight images of the Crab Nebula spanning the period from November 25, 2000 to April 6, 2001. The closeup image is 0.8 arcmin square. The other 7 images are 1.6 arcmin square. Courtesy NASA/CXC/ASU/J. Hester et al.
\label{f:crab}}
\end{center}
\end{figure}

The first X-rays focused by the telescope were observed on August 12, 1999. 
Figure~\ref{f:crab}\footnote{Pictures that are publicly available at
the \chandra\ web site at http://chandra.harvard.edu have credits labeled "Courtesy ... NASA/". The acronyms may be found at this site.} is a later example of one of the early images. 
The image of the Crab Nebula and its pulsar included a major new discovery
(Weisskopf et al. 2000) -
the bright inner elliptical ring showing the first direct observation of the
shock front where the wind of particles from the pulsar begins to radiate in
X-rays via the synchrotron process.

The Observatory's capability for high-resolution imaging has enabled detailed
studies of the structure of extended X-ray sources, including
supernova remnants, astrophysical jets, and hot gas in galaxies and clusters of
galaxies. 
Equally important have been \chandra's unique contributions to high-resolution
dispersive spectroscopy. 
The high spectral resolution of the \chandra\ gratings isolates individual lines
from the myriad of spectral lines, which would overlap at lower resolution.
The additional capability for spectrometric imaging allows studies of
structure, not only in X-ray intensity, but also in temperature and in chemical
composition. 
Through these observations, users have addressed and are continuing to address the most exciting topics in contemporary astrophysics.

In addition to mapping the structure of extended sources, the high angular 
resolution permits studies of discrete sources, which would otherwise be
impossible. 
From planetary systems to deep surveys of the faintest and most distant sources,
the scientific results from the first five years of \chandra\ operations have
been outstanding. 
We cannot possibly review all of the results, so in what follows we simply highlight a few.

\begin{figure}
\begin{center} 
\epsfysize=8cm
\epsfbox{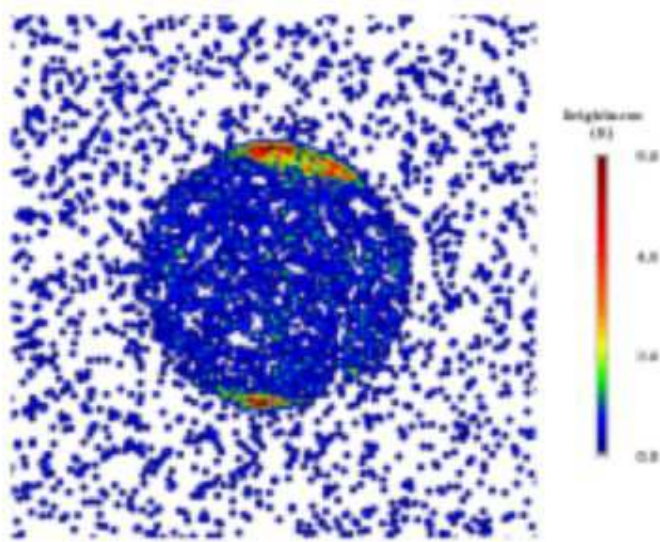} 
\caption{\chandra\ image of Jupiter showing the hot spots at high latitudes. The
image is 50--arcsec on a side. Courtesy R. Elsner.
\label{f:jupiter}}
\end{center}
\end{figure}

Figure~\ref{f:jupiter} shows hot spots at high (and unexpected) latitudes of the X-ray emission from the planet Jupiter that appear to pulsate at approximately a 45-minute period (Gladstone et al. 2002).
The X-rays are thought to be produced by particles bombarding the Jovian
atmosphere after precipitating along magnetic field lines. 
Several other planets have been shown to emit X-rays, albeit with different emission mechanisms. 
In the case of Mars, e.g. (Dennerl, 2002) fluorescent scattering of solar X-rays in the upper atmosphere produces the X-rays and the spectrum is dominated by a single narrow emission line, most likely caused by oxygen K-shell fluorescence.

\begin{figure}
\begin{center} 
\epsfysize=8cm
\epsfbox{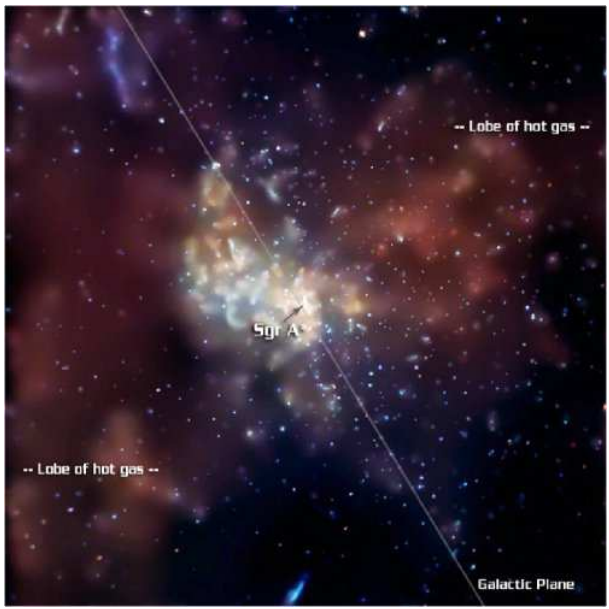} 
\caption{\chandra\ image of the Galactic Center. The image is 8.4 arcmin on a
side. Courtesy NASA/CXC/MIT/F. K. Baganoff et al.
\label{f:sgrA}}
\end{center}
\end{figure}

One of the most spectacular \chandra\ images is the one of the center of our own
galaxy (Baganoff et al. 2003) shown in Figure~\ref{f:sgrA} where one clearly sees both point-like discrete sources (over 1000)
and diffuse extended emission. 
The large amount of hot, X-ray-emitting gas has been heated and chemically
enriched by numerous stellar explosions.

The final legacy of \chandra\ may ultimately be led by the spectroscopic data.
The energy resolution, enabled by the quality of the optics and gratings, is providing new and extremely complex results. 
The broad bandpass of the grating spectrometers, combined with high resolution,
has proven equally important for astrophysical insights in situations where
spectral features are {\em not} present. 
For example, the remarkable line-free smoothness of the continuum of the
isolated neutron star RXJ~1856-3754 was a spectacular
surprise when revealed in detail using a 500~ks observation. 
This object is the nearest and brightest isolated neutron star candidate
(Walter, Wolk \& Neuh\"auser 1996), and it had been hoped that metal lines
formed in its outer atmosphere would provide a direct measurement of
gravitational redshift and insights into the equation of state of ultra-dense
matter.  
An early 55~ks exposure lacked the sensitivity to detect weak absorption
features (Burwitz et al.\ 2001), but the 500~ks LETG spectrum placed
stringent limits on the strengths of any absorption features (Drake et
al.\ 2002). The unexpected result has stimulated a number of theoretical investigations (e.g. Braje and Romani 2002, Tr\"umper et al. 2003, Mori and Ruderman 2003, Turolla et al. 2004) including speculations that the absence of features are due to the strong magnetic field or that its outer layers might lack an atmosphere and reside in a solid state or that the object is a slowly spinning magnetar.

Observations with the gratings are not only providing new astrophysical results,
they also provide a challenge to atomic physicists. 
The heart of the LETG+HRC-S bandpass covers the historically relatively
uncharted part of the soft X-ray spectrum from 25-70~\AA .  
Prior to \chandra, only a small handful of astrophysical observations had
been made at anything approaching high spectral resolution in this range: these
were of the solar corona using photographic spectrometers (Widing \& Sandlin
1986; Freeman \& Jones 1970; Schweizer \& Schmidtke 1971; Behring, Cohen, \&
Feldman 1972; Manson 1972; Acton et al. 1985).  
In comparison, LETG spectra of similar X-ray sources -- the coronae
of the solar-like stars $\alpha$~Centuri A (G2~V) and B (K1~V) and of Procyon
(Raassen et al.\ 2002, 2003) -- in this range are at the same time both daunting
and revealing.  
This spectral range contains a superposition of ``L-shell'' emission of abundant
elements such as Mg, Si, S and Ar, providing a challenge to spectroscopists
hoping to understand this region in terms of individual atomic transitions.
Drake et al. (2005) have shown that current radiative loss models in common
usage by X-ray astronomers underestimate the line flux in the 25-70~\AA\ range
by factors of up to 5.  
Laboratory efforts prompted by \chandra\ spectra, and the need for a better
theoretical description of plasma radiative emission in this spectral region,
are just beginning to unravel the tangle of lines into their parent ions (e.g.
Lepson et al. 2003 and references therein).

\begin{figure}
\begin{center} 
\epsfysize=8cm
\epsfbox{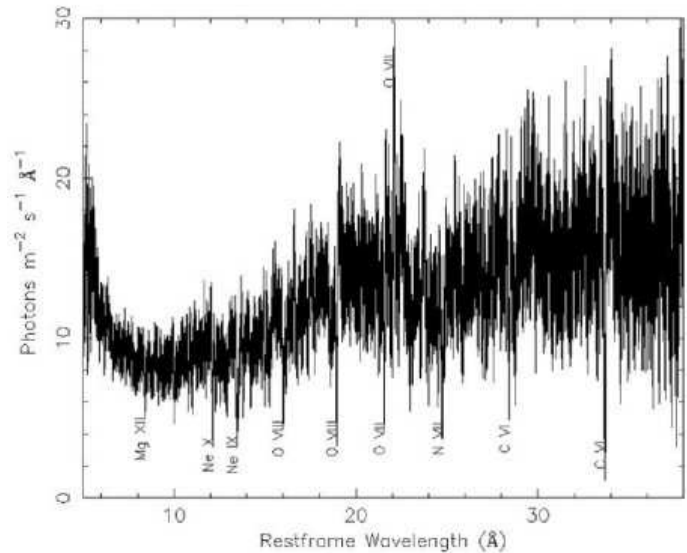} 
\caption{LETG/HRC spectrum of the Seyfert 1 galaxy NGC 5548 (Kaastra et al.
2000).
Several prominent absorption lines from H-like and He-like ions are marked, as
is the forbidden line of He-like oxygen.
\label{f:ngc5548}}
\end{center}
\end{figure}

\begin{figure}
\begin{center} 
\epsfysize=16cm
\epsfbox{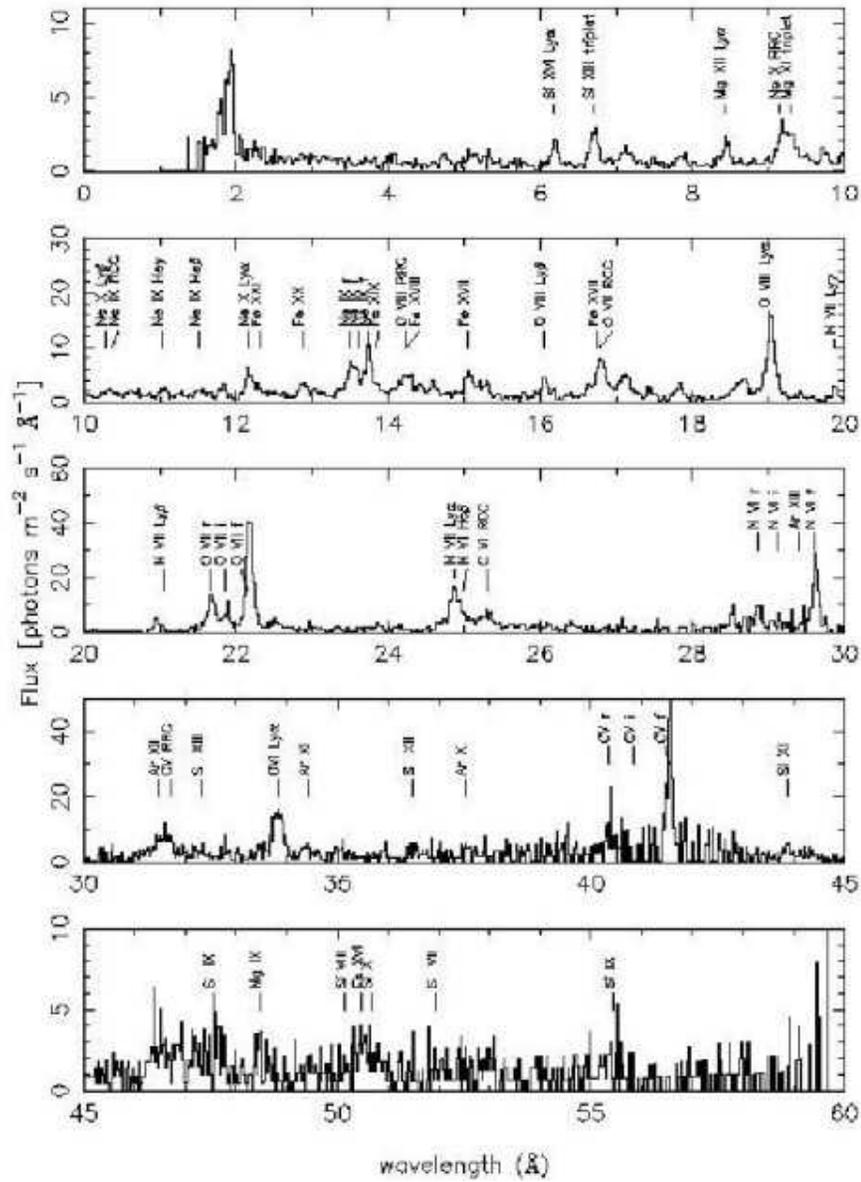} 
\caption{Emission-line spectrum from the Seyfert 2 galaxy NGC 1068. 
Kindly provided by A. Kinkhabwala.
\label{f:ngc1068}}
\end{center}
\end{figure}

High-resolution spectra of Seyfert galaxies are now providing new
details about the  physical and dynamical properties of material surrounding
the active nucleus.
For example, the Seyfert 1 active galaxy Mkn~478 was expected to exhibit
absorption lines at shorter wavelengths from a warm absorber that has
often been seen in the spectra of other Seyfert 1 galaxies, and emission lines
at wavelengths of $\sim 100$~\AA\ based on an analysis of EUVE spectra by Hwang
\& Bowyer (1997).  
Mkn~478 lies in a direction out of the galaxy that has a particularly low
neutral hydrogen column density, and so remains a strong source at these longer
wavelengths.  
Furthermore, for Seyfert-1s, whose signal is dominated by a bright X-ray
continuum from the central engine, the partially ionized circum-source material
introduces prominent patterns of absorption lines and edges. 
Figure~\ref{f:ngc5548}, e.g. shows a LETG/HRC-S spectrum of NGC 5548. 
This spectrum has dozens of absorption lines (Kaastra et al. 2000).
For Seyfert 2's the strong continuum from the central engine is not seen
directly, so the surrounding regions are seen in emission.
Figure~\ref{f:ngc1068} provides an example of a LETG/HRC observation of the
Seyfert 2, NGC 1068 (Brinkman et al. 2002).

\begin{figure}
\begin{center} 
\epsfysize=7cm
\epsfbox{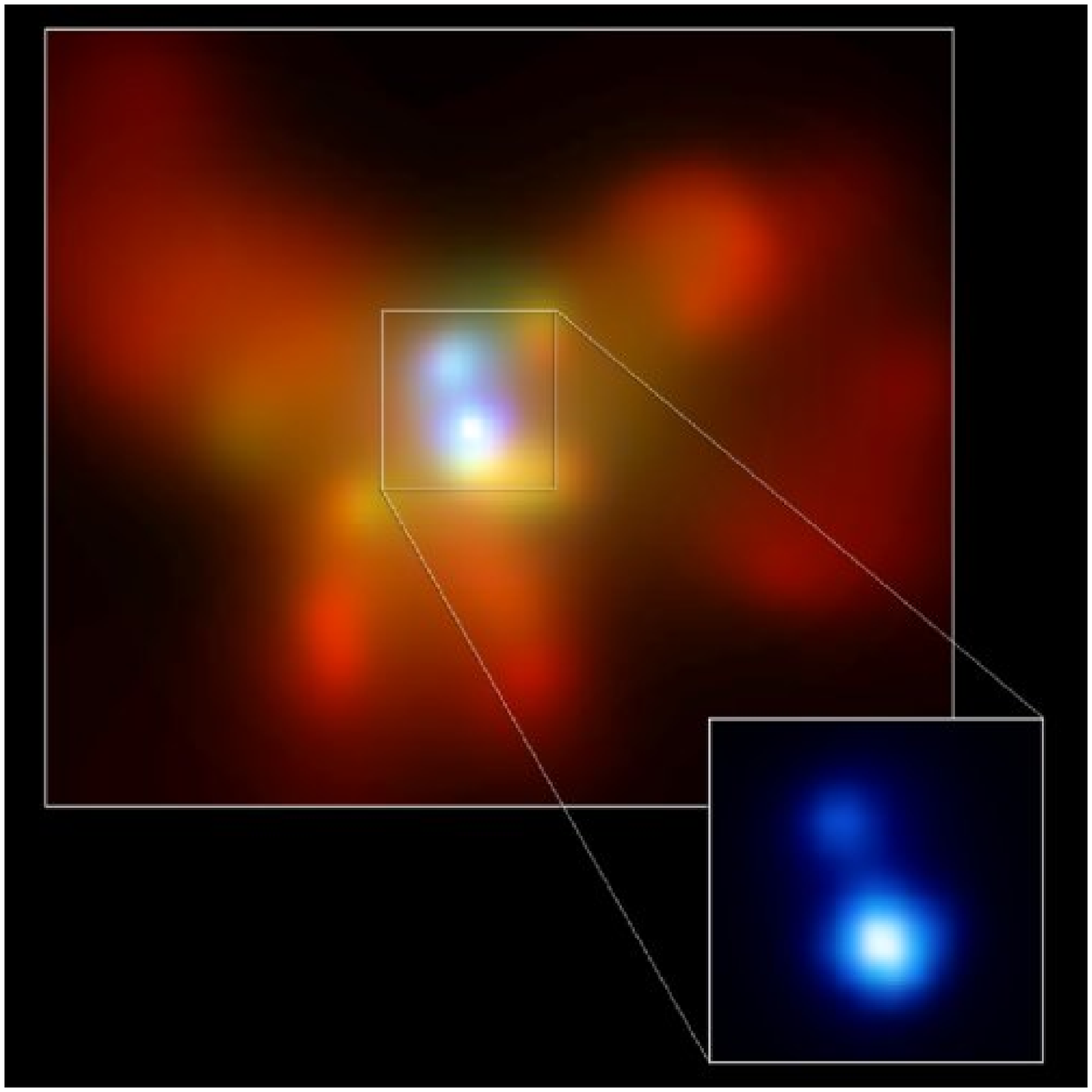} 
\caption{The \chandra\ image of NGC 6240, a butterfly-shaped galaxy that is the
product of the collision of two smaller galaxies, shows that the central region 
contains two active galactic nuclei. The image is 0.35 x 0.3 arcmin. Courtesy
NASA/CXC/MPE/S. Komossa et al.
\label{f:ngc6240}}
\end{center}
\end{figure}

\begin{figure}
\begin{center} 
\epsfysize=7cm
\epsfbox{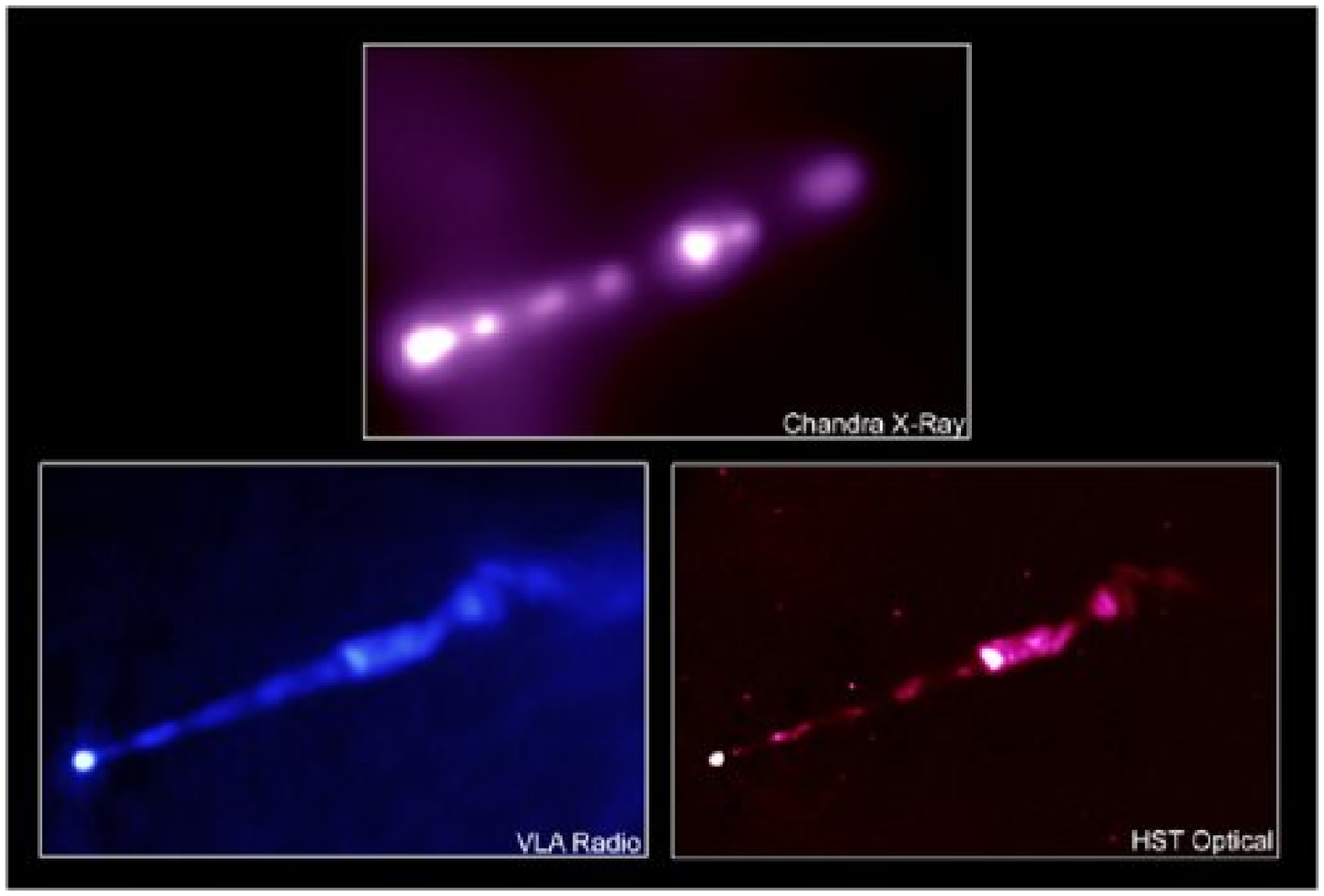} 
\caption{The X-ray jet emanating from the nucleus of the elliptical galaxy M87
as seen in three wavelength bands. Credits: X-ray: NASA/CXC/MIT/H. Marshall
et al. Radio: F. Zhou, F.Owen (NRAO), J.Biretta (STScI) Optical:
NASA/STScI/UMBC/E.Perlman et al. (2001).
\label{f:m87}}
\end{center}
\end{figure}

The study of active galaxies is one of the centerpieces of studies with the
Observatory. 
Figure~\ref{f:ngc6240} of NGC 6240 illustrates a recent spectacular result --- the first image of a double quasar nucleus (Komossa et al. 2003).
Figure~\ref{f:m87}\ of M87 illustrates multiwavelength observations of the jets from active galaxies. 
The \chandra\ X-ray image (Marshall et al. 2002) shows an irregular, knotty
structure similar to that seen at radio and optical (Perlman et al. 2001) wavelengths. 
However, the knots near the central core are much brighter in X-rays.
The jet phenomenon now appears to be ubiquitous in astronomical settings,
especially with regards to X-ray emission as e.g. the series of
observations of the outer jet of the Vela pulsar (Pavlov et al. 2003).

\begin{figure}
\begin{center} 
\epsfysize=6cm
\epsfbox{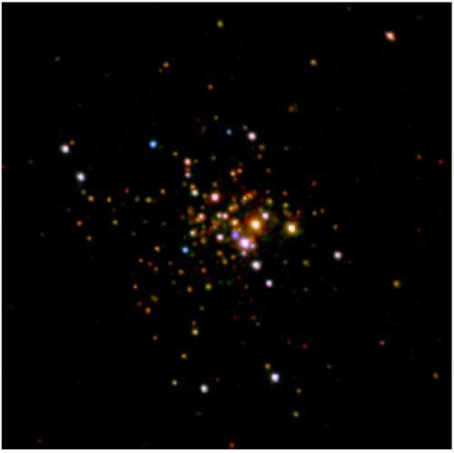} 
\caption{\chandra\ ACIS image of the globular cluster 47 Tucanae. The image 
covers the central 2.5' x 2.5'.
Courtesy NASA/CfA/J.Grindlay et al.
\label{f:47tuc}}
\end{center}
\end{figure}

\begin{figure}
\begin{center} 
\epsfysize=6cm
\epsfbox{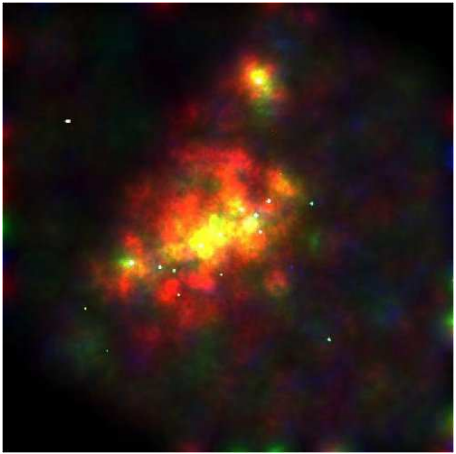} 
\caption{X-ray image of two interacting galaxies NGC 4490 and 4485. The image is
8-arcmin on a side. NGC 4490 is the larger of the two. Small dots indicate the
brightest X-ray sources. Courtesy Doug Swartz/USRA/MSFC. 
\label{f:ngc4490_4485}}
\end{center}
\end{figure}

One of the more important triumphs of the Observatory has been to use the
angular resolution and high sensitivity to perform detailed surveys of extended
objects such as globular clusters, galaxies, and clusters of galaxies. 
Figure~\ref{f:47tuc} shows one of the spectacular \chandra\ images of globular
clusters (Grindlay et al. 2001). 
A survey of two interacting galaxies is illustrated in
Figure~\ref{f:ngc4490_4485} where one sees emission from diffuse gas and bright
point sources.

A major triumph of \chandra\ (and XMM-Newton) high-resolution spectroscopic 
observations has been the discovery that the gas in the clusters is typically 
{\it not} cooling to below about 1-2 keV (see for example the discussion in
Fabian (2002) which indicates the presence of one (or more) heating
mechanisms).

\begin{figure}
\begin{center} 
\epsfysize=8cm
\epsfbox{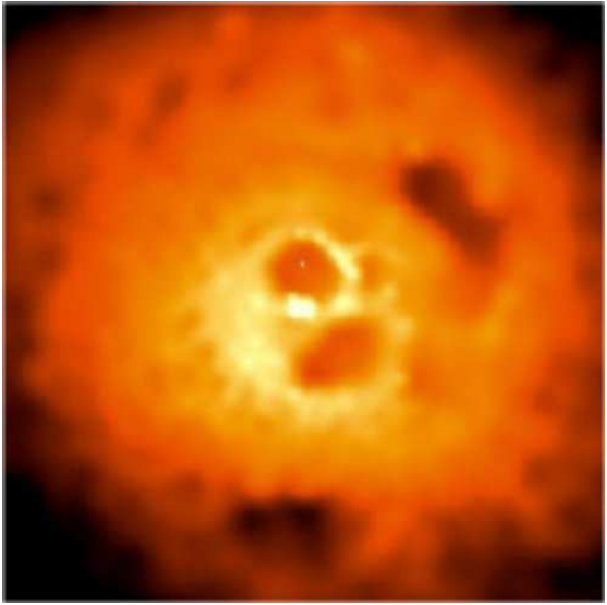} 
\caption{X-ray core of the Perseus  cluster. The image is about 3.5 arcmin on a
side. Courtesy NASA/IoA/A. Fabian et al.
\label{f:perseus}}
\end{center}
\end{figure}

\begin{figure}
\begin{center} 
\epsfysize=7cm
\epsfbox{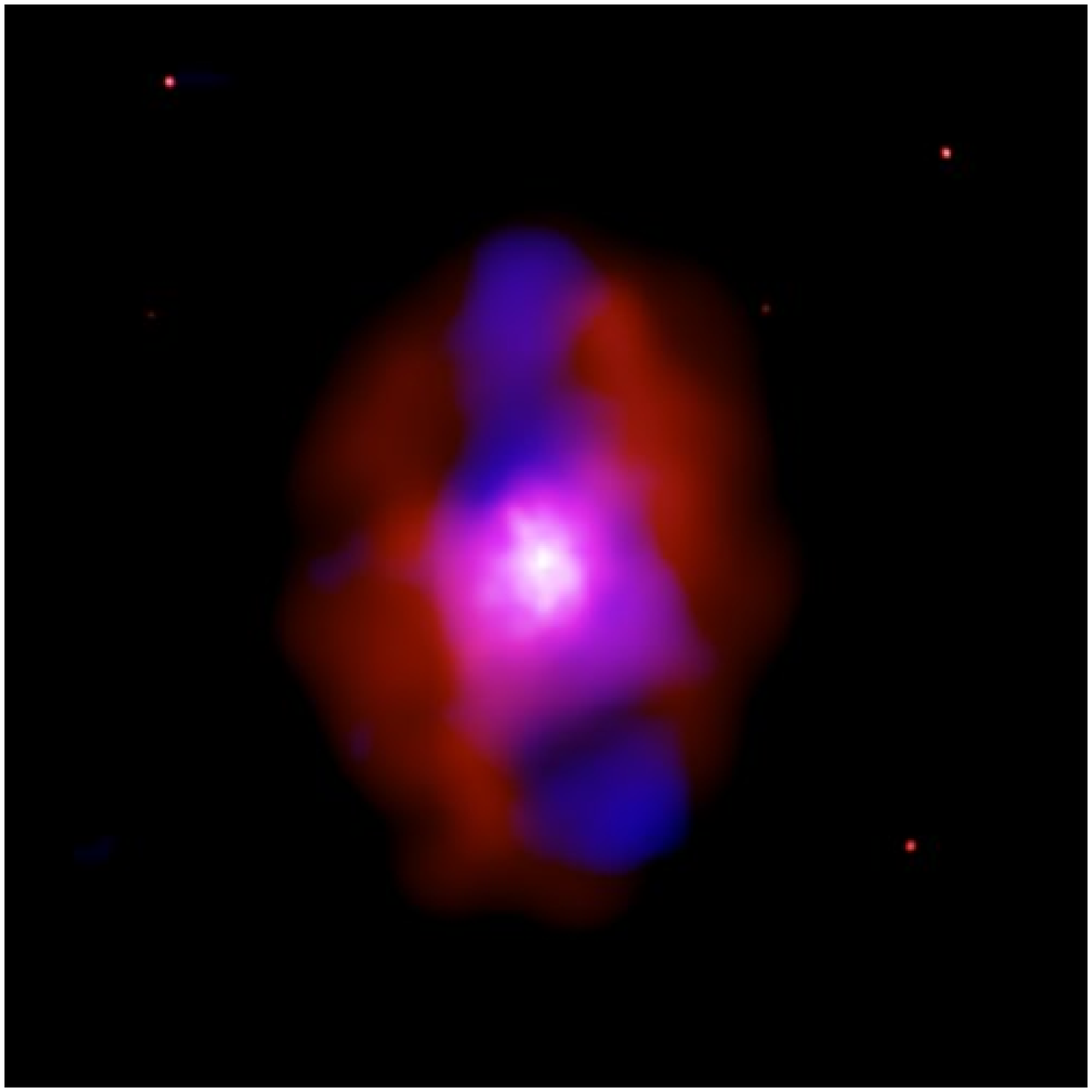} 
\caption{X-Ray and radio image of MS 0735.6+7421. The  image is  4.2' on a side. The X-Ray data are in red and the radio data are in blue. Courtesy NASA/CXC/Ohio U./B. McNamara.  
\label{f:ms0735}}
\end{center}
\end{figure}

\chandra\ observations of clusters of galaxies frequently exhibit previously
undetected structures with characteristic angular scales as small as a few arc
seconds. 
These structures include "bubbles" where there is strong radio emission, bow shocks, and cold fronts.  
Figure~\ref{f:perseus} of the Perseus cluster (Fabian et al. 2000) is an
example of bubbles produced in regions where there is strong radio emission.  
Figure~\ref{f:ms0735} is a spectacular recent example of giant cavities and shock fronts in an optically poor cluster, MS0735.6+7421, studied by McNamara et al. (2005). 
Radio lobes fill the X-ray faint cavities indicating that the hot gas is displaced and compressed by the expanding radio source. 
These authors find that the work required to inflate the cavity against the surrounding pressure is about $10^{61}$ ergs! 
Further consequences of this observation include the suggestion that the event was powered by accretion onto a central supermassive black hole. 
If so, the amount of material accreted in the last $10^8$ years was about $300 \times 10^8$ solar masses, about one third the total mass of the central black hole. 
 
\begin{figure}
\begin{center} 
\epsfysize=7cm
\epsfbox{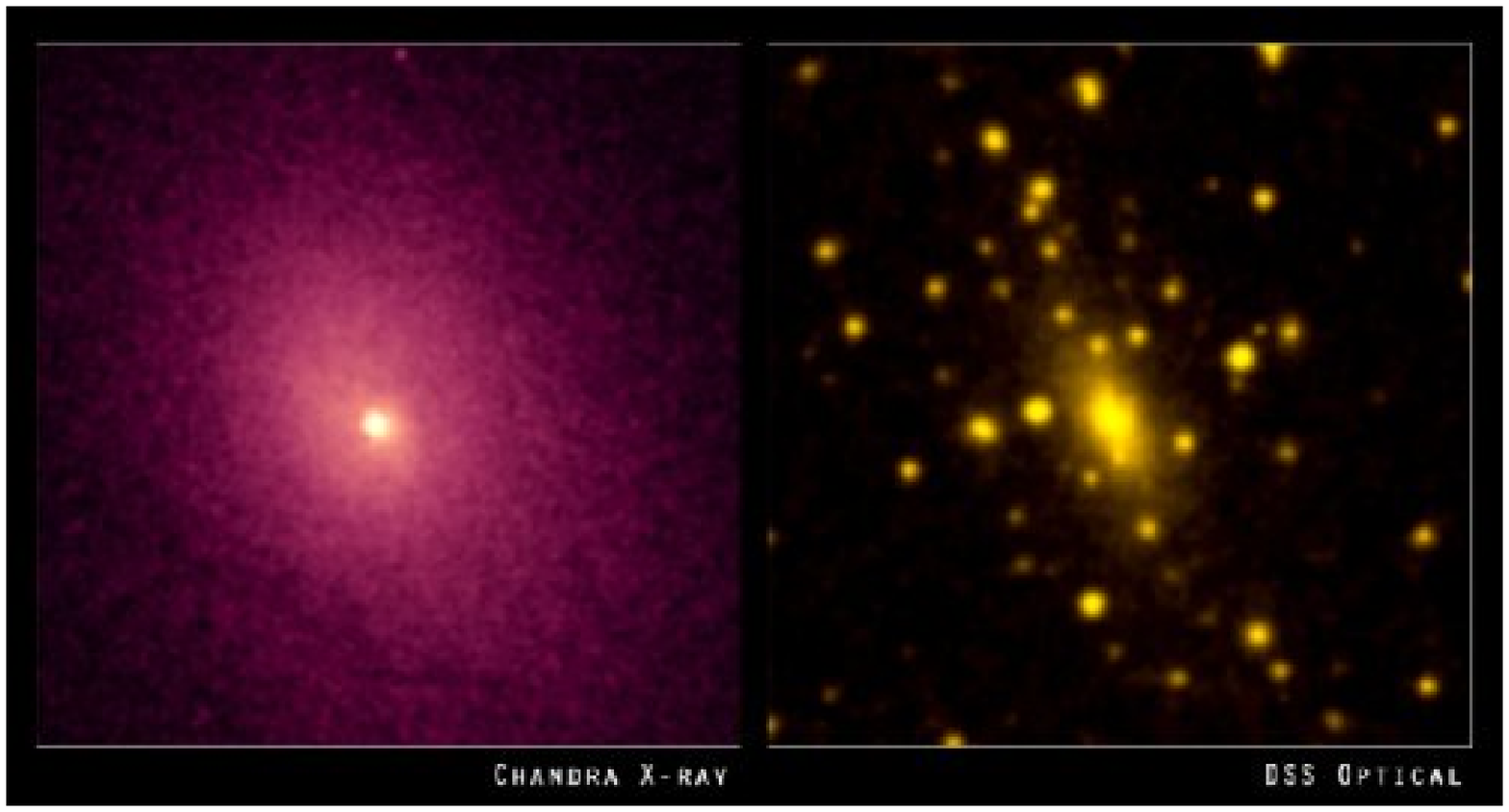} 
\caption{X-ray (left) and optical (right) images of the galaxy cluster Abell
2029. The images are 4-arcmin on a side. X-ray: NASA/CXC/UCI/A. Lewis et al.
Optical: Pal.Obs. DSS  
\label{f:abell2029}}
\end{center}
\end{figure}

Some clusters, such as Abell 2029 shown in Figure~\ref{f:abell2029}, do exhibit
a smoother relaxed structure.
Measurements of the temperature and density profiles of the gas, inwards toward
the central, dominant galaxy, provides a map of the gravitational potential,
and hence the location of the dark matter in the cluster.
The observers, Lewis, Buote, and Stocke (2003), showed that the dark
matter density increased toward the center in a manner consistent with cold
dark matter models. 

No discussion of data taken with the Observatory is complete without a mention
of the deep surveys. 
This work is an outgrowth of the study  of the diffuse X-ray background, the nature of which had been a puzzle for decades, although the lack of distortion
of the spectrum of the Cosmic Microwave Background placed a strong upper limit
to the possibility of a diffuse component (Mather et al. 1990).
Observations with ROSAT at energies below 2 keV made a major step in resolving a
significant fraction (70-80\%) into discrete objects (Hasinger et al. 1998).
Currently two long exposures have been accomplished - the \chandra\ Deep Field North (e.g. Alexander et al. 2003) with 2-Ms of exposure, and the \chandra\ deep field south (e.g. Giacconi et al. 2001) with 1-Msec. 
These surveys have extended the study of the background to flux levels more than
an order of magnitude fainter in the 0.5-2.0 keV band and have
resolved over 90\% of the background into a variety of discrete sources. 

Finally, we note \chandra's contributions to the study of dark energy. 
Studying the gas mass fraction in clusters of galaxies such as A2029, and assuming that this fraction is independent of redshift, Allen et al. (2004) detected the effects of dark energy on the distances to clusters of galaxies, independent of, but in confirmation of, previous such studies using Type 1a supernovae.  

\section{Conclusions and Acknowledgments}

Designed for a minimum of three years of operation, the \chandra\ X-ray
Observatory has now been operating successfully for more than five years.
The number of new discoveries has been legion and the Observatory has more
than lived up to its promise. 

This paper is in part a synopsis of a review paper "An Overview of the Performance of the \chandra\ X-Ray Observatory" that appeared in Experimental Astronomy, volume 16, pages 1-68 in 2003 by myself, T. Aldcroft, M. Bautz, R. Cameron, D. Dewey, J. Drake, C. Grant, H. Marshall, and S. Murray. I am indebted to my co-authors' contributions to that paper. 
I also acknowledge the major contributions to the success of the Observatory by the scientists and engineers associated with the instrument teams, the NASA Project at MSFC, the many contractors, and the CXC, with special thanks to its Director, Dr. H. Tananbaum.
Finally, I wish to acknowledge the tremendous contributions of Leon Van Speybroeck to the \chandra\ Project.

\end{document}